\documentstyle[12pt]{article}
\input psfig
\begin{document}
\begin{center}
{\bf High Energy Cosmic-Rays and Neutrinos 
from Cosmological Gamma-Ray Burst Fireballs}\\
\vspace*{1cm} 
{Eli Waxman\footnote{e-mail address: waxman@wicc.weizmann.ac.il}}\\
\vspace{0.15cm}
Dept. of Condensed Matter Physics, Weizmann Institute of Science \\
Rehovot 76100, Israel \\
\vspace*{1.5cm}  
{\bf Abstract} \\ \end{center}
\vspace*{5mm}

\noindent
The recent detection of delayed, low energy emission from
Gamma-Ray Burst (GRB) sources confirmed the 
cosmological origin of the bursts and provided support for models
where GRBs are produced by the dissipation
of the kinetic energy of relativistic fireballs.
In this review, ultra-high-energy, $>10^{19}$~eV, cosmic-ray and 
high energy, $\sim10^{14}$~eV, neutrino production in GRBs is discussed in the 
light of recent GRB and cosmic-ray observations.
Emphasis is put on model
predictions that can be tested with operating and planned 
cosmic-ray and neutrino detectors. 
The predicted neutrino intensity, $E_\nu^2 dN_\nu/
dE_\nu\sim3\times10^{-9}{\rm\ GeV/cm}^2{\rm s\ sr}$ for $10^{14}$~eV$<E_\nu<
10^{16}$~eV, implies that
a km$^2$ neutrino detector would observe tens of events
per year correlated with GRBs, and will be able
to test for neutrino properties with an
accuracy many orders of magnitude better than is currently possible. 
The predicted production rate of high-energy protons, 
which is consistent with that 
required to account for the observed  ultra-high-energy cosmic-ray (UHECR) 
flux, implies that operating and planned 
cosmic-ray detectors can test the GRB model for UHECR production.
If the predicted sources are found, cosmic-ray detectors
will provide us with a technique to
investigate the inter-galactic magnetic field.

\vspace*{1cm}
\begin{center}
Invited talk presented at the Nobel Symposium {\it Particle Physics and
The Universe}\\
Haga Slott, Sweden, August 1998 \\
\end{center} 

\newpage
\pagestyle{plain}

\section{Introduction}

The origin of Gamma-Ray Bursts (GRBs),
bursts of 0.1 MeV---1 MeV photons lasting for a few seconds, 
remained unknown for over 20 years, primarily because GRBs 
were not detected until the past year at wave-bands other than $\gamma$-rays 
(see \cite{Fishman} for review of $\gamma$-ray observations).
The isotropic distribution of bursts over the sky,
revealed by observations of the BATSE detector on board the Compton Gamma-Ray
Observatory, suggested that GRB sources lie at cosmological distances
\cite{BATSE92}. Adopting a cosmological distance scale to GRB sources, 
general phenomenological considerations were used to argue that the
bursts are produced by the dissipation of the kinetic
energy of a relativistic expanding fireball (see \cite{fireballs} for
reviews). 

The availability from the BeppoSAX satellite of accurate positions 
for GRBs shortly after their detection enabled during the past year
the detection of GRB ``afterglows,'' delayed X-ray \cite{Costa97}, optical 
\cite{Paradijs97} and radio \cite{Frail97} emission associated with GRB 
sources. Optical afterglow observations confirmed the cosmological 
origin of the bursts: Absorption lines detected in the afterglow of one
burst set a lower limit $z\ge0.835$ to its redshift \cite{z0508}, and 
the redshifts of three GRB host-galaxies were determined, 
GRB970508 at $z=0.835$ \cite{H0508}, GRB980703 at $z=0.965$ \cite{H0703}, 
and GRB971214 at $z=3.42$ \cite{H1214}. 
The characteristics of observed GRB afterglows,
the existence of which has been predicted by the fireball model
\cite{afterglow_predictions}, are broadly in agreement
with fireball model predictions and
therefore provide strong support for the model \cite{afterglow_consistent}. 
Furthermore, the detection \cite{Frail97} of predicted \cite{Goodman97} 
afterglow radio scintillation directly demonstrates the  
relativistic expansion of the source of GRB970508 \cite{WKF98}. 
It should be noted, however, that despite
the general success of the fireball model the 
underlying sources producing GRB fireballs remain unknown.

Much like the underlying GRB source, the origin of ultra-high energy
cosmic-rays (UHECRs), cosmic-rays of
energy $>10^{19}$~eV, is unknown (see \cite{Cronin96} for a
recent review). Most of the sources of cosmic-rays that have been
proposed have difficulties in accelerating particles up to the highest
observed energy \cite{Hillas84,Cronin96}, which is in excess of $10^{20}$~eV 
\cite{Watson91,Yakutsk,Bird934,Hayashida945,Takeda98}.
Furthermore, since the distance traveled by the highest energy 
particles must be smaller, due to interaction with radiation backgrounds 
\cite{GZK}, than 100~Mpc \cite{Aharonian94}, their arrival 
directions are inconsistent with the position of any astrophysical object 
that is likely to produce high energy particles \cite{CR_objects}.
Well before the recent confirmation of the hypothesis that
GRBs are of cosmological origin, it has been shown that cosmological 
fireballs are likely sources of $>10^{19}$~eV protons. 
The physical conditions in the 
fireball dissipation region imply that protons may be Fermi accelerated in 
this region to energy $>10^{20}{\rm eV}$ \cite{W95a,Vietri95}. 
In addition, the average rate at which energy is emitted as $\gamma$-rays
by GRBs is remarkably 
comparable to the energy generation rate of UHECRs in a model where
UHECRs are produced by a cosmological distribution of sources 
\cite{W95a,W95b}. 
These two facts suggest that GRBs and UHECRs have
a common origin\footnote{Milgrom \& Usov suggested \cite{MnU95}
a GRB--UHECR association based on the overlap of the Fly's Eye highest 
energy cosmic ray arrival direction error box with the position error box
of a bright GRB. A correlation between GRB and UHECR arrival 
directions is not expected, as explained in \S4, for cosmological GRBs.}. 
The GRB fireball model for UHECR production makes several 
unique predictions \cite{MnW96,WnM96}, which can be tested with
operating (HiRes \cite{HiRes}), and planned (Auger \cite{Auger},
Telescope-Array \cite{TA}) UHECR detectors. Possibly the most interesting
consequence of proton acceleration in GRB fireballs
is the conversion of a significant fraction, $\simeq10\%$,
of the fireball energy to an accompanying burst of $\sim10^{14}{\rm eV}$ 
neutrinos \cite{WnB97}. The predicted flux implies detection of
tens of events per year correlated with GRBs in 
planned km$^2$ neutrino detectors
(AMANDA-II and DeepIce \cite{Barwick_Nobel} extensions of the operating
AMANDA detector \cite{AMANDA},
ANTARES \cite{ANTARES}, NESTOR \cite{NESTOR}).

In this paper,
high-energy cosmic-ray and neutrino production in GRBs is discussed in the 
light of recent GRB and ultra-high-energy cosmic-ray observations. 
In \S2 the fireball model is briefly described, 
and implications of recent afterglow
observations are discussed, which are of importance for high energy particle 
production. A more detailed discussion of GRBs and the 
fireball model is given in a 
separate contribution to these proceedings \cite{Rees_Nobel}.
In section \S3 recent UHECR observations,
described in detail elsewhere in these proceedings \cite{Watson_Nobel},
are discussed. The flux and spectrum measured by the Fly's Eye, Yakutsk,
and AGASA experiments are compared with the prediction of a model where
UHECRs are protons accelerated to high energy by Fermi shock
acceleration in sources which are uniformly distributed in the
universe. The main spectral feature predicted by such a model
is a ``GZK cutoff'' \cite{GZK}, a suppression of 
UHECR flux above $\sim5\times10^{19}$~eV due to interaction of protons 
with the microwave background. We show that present data do not allow to 
confirm or rule
out the existence of the predicted suppression. While both Fly's Eye and 
Yakutsk data show a suppression at $2\sigma$ significance, 
a discrepancy may be emerging between this data and the results of the AGASA 
experiment: the Fly's Eye experiment and the Yakutsk experiment each report 
one event beyond $10^{20}$~eV (close to the
average number of 1.5 events predicted by the cosmological model), 
while the AGASA experiment reports 6 events 
for similar exposure. Much larger
exposure than presently available is required to determine whether or not
a GZK ``cutoff'' exists.

The production in GRB fireballs of UHECRs is discussed in \S4. Predictions
of the GRB model for UHECR production, that can be tested with future 
UHECR experiments, are discussed in \S5. 
High energy neutrino production in fireballs and its implications for future
high energy neutrino detectors are discussed in \S6. 
The discussion in \S4--\S6 of UHECR and neutrino production in GRBs
is similar to the analysis presented prior to the discovery of GRB afterglow
\cite{W95a,WnB97}. 
Some quantitative modifications are introduced due to the revised GRB energy
scale, $\sim10^{53}$~erg (for isotropic emission) implied by afterglow
observations compared to $\sim10^{52}$~erg previously assumed, and due to 
some, yet inconclusive, evidence that the local GRB rate is lower
than previously estimated, $\sim1{\rm\ Gpc}^{-3}{\rm yr}^{-1}$ compared
to $\sim10{\rm\ Gpc}^{-3}{\rm yr}^{-1}$.
We also address some criticism of the GRB model of UHECR production 
recently made in the literature regarding 
the energy loss of protons escaping the fireball
\cite{RnM_adiabatic}, and the acceleration process \cite{Gallant98}. We
show that this criticism is inapplicable to the model for UHECR 
production discussed here, which was proposed in \cite{W95a}.
Finally, in \S7 a summary is presented of
the main points discussed in the paper.

\section{GRB fireballs and afterglow observations}

\subsection{The fireball model}

General phenomenological considerations, based on $\gamma$-ray
observations, indicate that,
regardless of the nature of the underlying sources, 
GRBs are produced by the dissipation of the kinetic energy of a 
relativistic expanding fireball. 
The rapid rise time and short duration, $\sim1$~ms, 
observed in some bursts \cite{Bhat92_Fishman94} imply
that the sources are compact, with a linear scale $r_0\sim10^7$~cm. The
high $\gamma$-ray luminosity required for cosmological bursts, 
$L_\gamma\sim10^{52}{\rm erg\ s}^{-1}$,
then results in an initially optically thick (to pair creation) plasma
of photons, electrons and positrons, which expands and accelerates to 
relativistic velocities \cite{Bohdan_Goodman86}. 
This is true whether the energy is released instantaneously, 
i.e. over a time scale $r_0/c$,
or as a wind over a duration comparable to the entire
burst duration ($\sim$seconds). In fact, the hardness of the observed
photon spectra, which extends to $\sim100$~MeV, implies 
that the $\gamma$-ray emitting region must be moving relativistically, 
with a Lorentz factor $\Gamma\sim300$ \cite{Krolik_Baring}, 
in order that
the fireball pair-production optical depth be small for the observed 
high energy photons.

If the observed radiation is due
to photons escaping the fireball/wind as it becomes optically thin, two 
problems arise. First, the photon spectrum is quasi-thermal,
in contrast with observations. Second, 
the source size, $r_0\sim10^7$~cm, and the total energy emitted in gamma-rays,
$\sim10^{53}$~erg, suggests that the underlying energy source is related
to the gravitational collapse of $\sim 1 M_\odot$ object.
Thus, the plasma is expected to be ``loaded''
with baryons which may be injected with the radiation or present in the 
atmosphere surrounding the source. A small baryonic load, $\geq10^{-8}
{M_\odot}$, increases the optical depth (due to Thomson scattering) so that 
most of the radiation energy is converted to
kinetic energy of the relativistically expanding baryons before the plasma
becomes optically thin \cite{kinetic}. 
To overcome both problems it was proposed \cite{RnM92} that the
observed burst is produced once the kinetic energy of the ultra-relativistic 
ejecta is re-randomized by some dissipation process at large radius, beyond
the Thomson photosphere, and then radiated as $\gamma$-rays. Collision 
of the relativistic baryons
with the inter-stellar medium \cite{RnM92}, and 
internal collisions within the ejecta
itself \cite{internal}, 
were proposed as possible dissipation processes.

Most GRBs show variability on time scales much shorter than (typically
one hundredth of) the total
GRB duration. Such variability is hard to explain in models where the
energy dissipation is due to external shocks \cite{Woods95,SnP_var}.
Thus, it is believed that internal collisions are responsible for the
emission of gamma-rays. At small radius, 
the fireball bulk Lorentz factor, $\Gamma$, 
grows linearly with radius, until most of the wind energy is converted
to kinetic energy and $\Gamma$ saturates at $\Gamma\sim300$.
Variability of the source on time scale $\Delta t$, resulting
in fluctuations in the wind bulk Lorentz factor $\Gamma$ on similar
time scale, then leads to internal shocks in the expanding fireball
at a radius
\begin{equation}
r_i\approx\Gamma^2c\Delta t=3\times10^{13}\Gamma^2_{300}\Delta t_{\rm10ms}
{\rm\ cm},
\label{rd}
\end{equation}
where $\Gamma=300\Gamma_{300}$, $\Delta t=10\Delta t_{\rm10ms}$~ms.
If the Lorentz factor variability within the wind is significant,
internal shocks would reconvert a substantial 
part of the kinetic energy to internal energy. It is assumed that
this energy is then radiated as 
$\gamma$-rays by synchrotron and inverse-Compton emission of
shock-accelerated electrons. For internal collisions, the observed
gamma-ray variability time, $\sim r_i/\Gamma^2 c\approx\Delta t$,
reflects the variability time of the underlying source, and the GRB
duration reflects the duration over which energy is emitted from the
source. Since the wind Lorentz factor is expected to fluctuate on
time scales ranging from the shortest variability time $\Delta t$ to the
wind duration $T$, internal collisions will take place over a range
of radii, $r\sim r_i=\Gamma^2c\Delta t$ to $r\sim\Gamma^2cT$. 
A large fraction of bursts detected by BATSE show variability
on the shortest resolved time scale, $\sim10$~ms. Our choice in 
Eq. (\ref{rd}) of $\Delta t=10$~ms as a representative
value is therefore conservative, in the sense that 
the shortest variability time may be
significantly smaller. In fact, recent analysis indicates that variability
on $\sim1$~ms is common \cite{Schaefer98}.

Internal shocks are expected to be ``mildly'' relativistic in the fireball 
rest frame \cite{internal}, i.e. characterized by Lorentz factor 
$\gamma_i-1\sim1$ (since adjacent shells within the wind are expected to
expand with Lorentz factors which do not differ by more than an
order of magnitude). 
The internal shocks would therefore heat the protons
to random velocities (in the wind frame) $\gamma_p-1\sim1$. The characteristic
frequency of synchrotron emission is determined by the characteristic energy
of the electrons and by the strength of the magnetic field. These are
determined by assuming that the fraction of energy carried
by electrons is $\xi_e$, implying a characteristic rest frame electron Lorentz
factor $\gamma_e=\xi_e(m_p/m_e)$, and that a fraction $\xi_B$ of the energy 
is carried by the magnetic field, implying 
$4\pi r_i^2c\Gamma^2B^2/8\pi=\xi_B L$ where $L$ is the 
total wind luminosity. Since the electron synchrotron cooling time is short
compared to the wind expansion time, electrons lose their energy
radiatively and $L\approx L_\gamma/\xi_e$. 
The characteristic observed energy of synchrotron photons, 
$E_\gamma=\Gamma\hbar\gamma_e^2 eB/m_ec$, is therefore
\begin{equation}
E_\gamma\approx0.1(\xi_B/0.3)^{1/2}(\xi_e/0.3)^{3/2}{L_{\gamma,52}^{1/2}
\over\Gamma_{300}^2\Delta t_{\rm10ms}}{\rm MeV}, 
\label{Eg}
\end{equation}
where $L_\gamma=10^{52}L_{\gamma,52}{\rm erg/s}$.
At present, there is no theory that allows the determination of 
the values of the equipartition fractions $\xi_e$ and $\xi_B$. However,
it is encouraging that for values close to equipartition, the photon 
energy predicted by the model is similar to that observed.

As the fireball expands, it drives a relativistic shock (blastwave)
into the surrounding
gas, e.g. into the inter-stellar medium (ISM) gas if the explosion 
occurs within a galaxy. In what follows, we refer to the surrounding gas
as ``ISM gas,'' although the gas need not necessarily be inter-stellar.
At early time, the fireball is little affected by the interaction with the 
ISM. At late time, most of the fireball energy is transferred to the ISM, and
the flow approaches the self-similar blast-wave solution of Blandford \&
McKee \cite{BnM76}. At this stage a single shock propagates into the ISM,
behind which the gas expands with Lorentz factor
\begin{equation}
\Gamma_{BM}(r)=150\left({E_{53}\over n_1}\right)^{1/2}r_{17}^{-3/2}
\label{Gamma}
\end{equation}
where $E=10^{53}E_{53}$~erg is the (isotropic) fireball energy,
$n=1n_1{\rm\ cm}^{-3}$ is the ISM number density, and $r=10^{17}r_{17}$~cm
is the shell radius. The expansion becomes self-similar at a radius $r$
where two conditions are met: the Lorentz factor $\Gamma_{BM}(r)$ inferred 
from the self-similar solution is smaller than the initial Lorentz factor 
$\Gamma$, and the width of the shell into which the shocked ISM is 
compressed in the self-similar solution, 
$\approx r/10\Gamma_{BM}^2(r)$, is larger than the initial 
fireball shell width $cT$. The first and second conditions are met for
$r>r_\Gamma\equiv(17E/16\pi\Gamma^2 n m_p c^2)^{1/3}$ and
$r>r_T\equiv(10\times17EcT/16\pi n m_p c^2)^{1/4}$ respectively. Thus,
the transition to self-similar, external-shock flow occurs at
\begin{equation}
r_e=5.2\times10^{16}\left({E_{53}\over n_1}\right)^{1/4}
\max\left[1.0T^{1/4}_{1},1.2\left({E_{53}\over n_1}\right)^{1/12}
\Gamma_{300}^{-2/3}\right]\quad{\rm cm},
\label{re}
\end{equation}
where $T=1T_{1}$~s. 

Internal, mildly-relativistic shocks within the fireball shell result both 
from variability of the source, at $r_i<r<r_e$, and from the
interaction of the fireball with the surrounding gas, during the
transition to a self-similar expansion at $r\sim r_e$, where
reverse shocks propagate into the expanding fireball ejecta
and decelerate it. From
Eq. (\ref{re}) we infer that for typical fireball parameters the two
conditions required for transition to self-similar external shock flow 
are satisfied at a similar radius, $r_e\approx r_\Gamma\approx r_T$.
This implies that significant deceleration of the fireball shell 
does not take place prior to the transition to self-similar behavior. 
This, in
turn, implies that the reverse shocks propagating into fireball
ejecta are only mildly relativistic. Thus,
the characteristics of internal shocks due to interaction
with surrounding gas are similar to those of internal shocks due to
variability on time scale $\sim T$. In the discussion that follows
we therefore do not discuss the reverse shocks separately from the 
internal shocks.

The shock driven into the ISM continuously heats new gas, and
produces relativistic 
electrons that may produce the delayed radiation, ``afterglow'',
observed on time scales of
days to months. As the shock-wave decelerates, the emission shifts to
lower frequency with time. Since proton acceleration to high energy takes
place only in the internal shocks, $r\le r_e$ (see \S4), we do
not discuss further the theory of afterglow emission.

\subsection{Afterglow observations}

Afterglow observations confirmed, as discussed in the Introduction, 
the cosmological origin of GRBs, and are consistent with delayed GRB emission
being synchrotron radiation of electrons accelerated to
high energy in the highly relativistic shock driven by the fireball into
its surrounding gas. Since we are interested mainly in the earlier, internal
collision phase, of fireball evolution, we do not discuss afterglow 
observations in detail. We note, however, two implications of
afterglow observations which are of importance for the discussion of UHECR
production. 

The first implication is related to the GRB energy scale. The gamma-ray
energy emitted by the three GRBs with measured redshifts in the energy
range of 20~keV to 2~MeV (assuming spherical symmetry) is 
$\approx10^{52}$~erg, $\approx0.8\times10^{53}$~erg and 
$\approx3\times10^{53}$~erg
for GRB970508 ($z=0.835$), GRB980703 ($z=0.966$) and GRB971214 ($z=3.42$)
respectively (here, and throughout the paper, we assume an open universe, 
$\Omega=0.2$, $\Lambda=0$, and $H_0=75{\rm\ km/s\ Mpc}$). 
This implies that GRBs are not ``standard candles,'' and
that the characteristic gamma-ray energy (luminosity) is 
$\sim10^{53}$~erg ($\sim10^{52}{\rm\ erg/s}$) rather than
$\sim10^{52}$~erg ($\sim10^{51}{\rm\ erg/s}$) as commonly assumed in the
past. 
Performing a detailed analysis Mao \& Mo
(1998) find, for example, that the typical luminosity of observed GRBs
is $\sim10^{52}{\rm\ erg/s}$ (Note that Mao \& Mo use 
$H_0=100{\rm\ km/Mpc\ s}$ and quote
luminosities and energies in the 50--300~keV band only).

The second implication relates to the GRB rate.
Krumholtz, Thorsett \& Harrison \cite{KTH98}
(see also \cite{HnF98}) have demonstrated that, based on the data, it
is impossible to distinguish between models where the GRB rate per
unit comoving volume is independent of redshift, and models where it
evolves rapidly, e.g. following star formation rate (previous claims to the 
contrary were based on the assumption, now known to be invalid, 
that GRBs are standard candles). Most observed GRBs originate at 
the redshift range $z\sim1$ to $z\sim2$ \cite{KTH98,HnF98}, and the observed
GRB rate essentially determines the GRB rate per unit volume at that redshift.
The present rate is less well constrained and ranges from 
$R_{\rm GRB}\sim1/{\rm Gpc}^3{\rm yr}$, assuming the GRB rate evolves
rapidly as the star-formation rate, to 
$R_{\rm GRB}\sim10/{\rm Gpc}^3{\rm yr}$, assuming the GRB rate is
independent of redshift. There is some evidence supporting the 
hypothesis that the GRB rate follows the star formation rate \cite{GRB_SFR}.
The evidence is, however, not yet conclusive.

\section{UHECR observations and their implications}

Fly's Eye \cite{Bird934} and AGASA \cite{Hayashida945,Takeda98} results
confirm the flattening of the cosmic-ray spectrum at $\sim10^{19}$~eV,
evidence for which existed in previous experiments with weaker statistics
\cite{Watson91}. Fly's Eye data is well fitted in the energy range 
$10^{17.6}$~eV to $10^{19.6}$~eV by a sum of two power laws: A
steeper component, with differential number spectrum
$J\propto E^{-3.50}$, dominating at lower
energy, and a shallower component, $J\propto E^{-2.61}$, 
dominating at higher energy, $E>10^{19}$~eV.
The flattening of the spectrum, combined with the lack of anisotropy 
and the evidence for a change in composition from heavy nuclei at low
energy to light nuclei (protons) at high energy \cite{Watson91,Bird934},
suggest that an extra-Galactic source of protons dominates the flux at
high energy.

\begin{figure}
\centerline{\psfig{figure=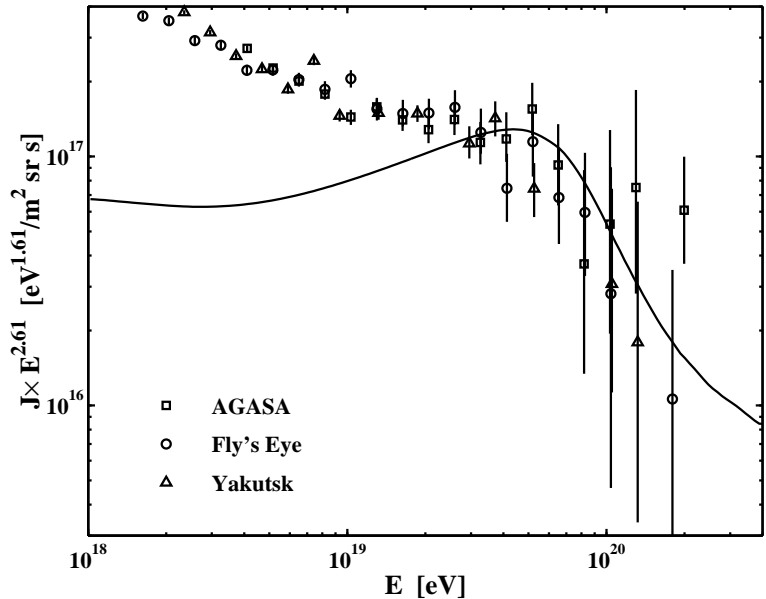,width=4.in}}
\caption{
The UHECR flux expected in a cosmological model, where high-energy protons 
are produced at a rate $E^2 d\dot n_{CR}/dE=0.8\times10^{44}
{\rm erg/Mpc}^3{\rm yr}$, compared to the Fly's Eye, Yakutsk and AGASA data. 
$1\sigma$ flux error bars are shown. The highest energy points are derived
assuming the detected events (1 for Fly's Eye and Yakutsk, 
4 for AGASA) represent a
uniform flux over the energy range $10^{20}$~eV--$3\times10^{20}$~eV.}
\label{fig1}
\end{figure}

In Fig. 1 we compare the UHECR spectrum,
reported by the Fly's Eye, the Yakutsk, 
and the AGASA experiments \cite{Bird934,Yakutsk,Takeda98}, 
with that expected from 
a homogeneous cosmological distribution of sources, each generating
a power law differential spectrum of high energy protons
$dN/dE\propto E^{-2}$. This spectrum is expected for Fermi shock acceleration 
\cite{Bell_plus} (see \S4.1 for discussion of the GRB UHECR model). 
The absolute flux measured at
$3\times10^{18}$~eV differs between the various experiments,
corresponding to a systematic $\simeq10\%$ ($\simeq20\%$) over-estimate 
of event energies in the AGASA (Yakutsk)
experiment compared to the Fly's Eye experiment (see also \cite{Hayashida945}).
In Fig. 1, the Yakutsk energy normalization is used. 
For the model calculation, an open universe, $\Omega=0.2$, $\Lambda=0$ and
$H_0=75{\rm km/ Mpc\ s}$ were assumed. The calculation is similar
to that described in \cite{W95b}. The generation rate of cosmic-rays 
(per unit comoving volume)
was assumed to evolve rapidly with redshift following
the luminosity density evolution of QSOs \cite{QSO}, which is 
also similar to that describing the evolution of star formation rate 
\cite{SFR}: $\dot n_{CR}(z)\propto(1+z)^{\alpha}$ with $\alpha\approx3$ 
\cite{QSOl} at low redshift, 
$z<1.9$, $\dot n_{CR}(z)={\rm Const.}$ for $1.9<z<2.7$, 
and an exponential decay at 
$z>2.7$ \cite{QSOh}. The cosmic-ray spectrum at energy $>10^{19}$~eV is 
little affected by modifications of the cosmological parameters or of
the redshift evolution of cosmic-ray generation rate. This is due to the
fact that cosmic-rays at this energy originate from distances shorter than
several hundred Mpc. The spectrum and flux at $E>10^{19}$~eV is mainly 
determined by the present ($z=0$) generation rate and spectrum, which
in the model shown in Fig. 1 is $E^2 (d\dot n_{CR}/dE)_{z=0}=0.8\times10^{44}
{\rm erg/Mpc}^3{\rm yr}$.

The suppression of model flux above $\sim10^{19.7}$~eV, compared to a
power-law extrapolation of the flux at low energy, is  
due to energy loss of high energy protons
in interaction with the microwave background, i.e. to the ``GZK cutoff''
\cite{GZK}. This is the characteristic signature of cosmological
source distribution. It is clear from Fig. 1 that present data does not
allow to confirm or rule out the existence of the ``cutoff'' with high 
confidence. Nevertheless, some evidence for the cutoff does exit. 
Both Fly's Eye and Yakutsk data show a deficit in the number of events
detected above $10^{19.7}$~eV, compared to the number expected 
based on extrapolation of the Fly's Eye shallower power-law fit 
for $E<10^{19.6}$~eV, $J\propto E^{-2.61}$. The deficit is, however, only
at a $2\sigma$ confidence level \cite{W95a}. The AGASA data is consistent 
with Fly's Eye and Yakutsk results below $10^{20}$~eV.
A discrepancy may be emerging, see Fig. 1, 
at higher energy, $>10^{20}$~eV, 
where the Fly's Eye and Yakutsk experiments detect 1 event each,
and the AGASA experiment detects 6 events for similar exposure (a
$\sim2\sigma$ discrepancy). 

We therefore conclude, that a scenario where UHECRs are produced by
a cosmological distribution of sources, generating
high energy protons at a rate 
$E^2 (d\dot n_{CR}/dE)_{z=0}\approx10^{44}{\rm erg/Mpc}^3{\rm yr}$, 
is consistent with the observed flux and 
spectrum of cosmic-rays in the energy range of
$10^{19}$~eV to $10^{20}$~eV. The flux predicted by this model above
$10^{20}$~eV is consistent with that measured by the Fly's Eye and
the Yakutsk experiments, while AGASA results suggest a higher flux at this
energy. The statistical significance of the discrepancy between
the experiments (or between the AGASA results and model prediction
above $10^{20}$~eV) is not high. Clearly, much larger exposure than
presently available is required to accurately determine the UHECR
spectrum and flux above $5\times10^{19}$~eV.

\section{UHECRs from GRB fireballs}

\subsection{Fermi acceleration in GRBs}

In the fireball model, the observed radiation is produced, both during
the GRB and the afterglow, by synchrotron emission of shock accelerated
electrons. In the region where electrons are accelerated, 
protons are also expected to be
shock accelerated. This is similar to what is thought to occur in supernovae 
remnant shocks, where synchrotron radiation of accelerated electrons is the
likely source of non-thermal X-rays (recent ASCA observations give evidence
for acceleration of electrons in the remnant of SN1006 to $10^{14}{\rm eV}$ 
\cite{SN1006}), and where shock acceleration of protons is believed to
produce cosmic rays with energy extending to $\sim10^{15}{\rm eV}$ (see, e.g.,
\cite{Bland87} for review). Thus, it is likely that protons, as well
as electrons, are accelerated to high energy within GRB fireballs.
Let us consider the constraints that should be satisfied
by the fireball parameters in order to allow acceleration of protons to $\sim
10^{20}$~eV.

We consider proton Fermi acceleration in fireball internal shocks, which
take place as the fireball expands over a range of radii
$r_i\sim10^{14}{\rm\ cm}\le r\le r_e\sim10^{16}{\rm\ cm}$ 
[cf. Eqs. (\ref{rd},\ref{re})].
As mentioned in \S2.1, internal shocks are due to variability of the source,
on time scales ranging from the shortest variability time $\Delta t\sim10$~ms 
to the wind duration $T\sim10$~s. In addition, internal shocks arise also
due to interaction with ambient  
gas. The characteristics of the latter internal shocks, 
which occur at $r\sim r_e$ during the transition to
self-similar expansion, are similar to those of internal 
collisions due to variability on time scale $T$ [see discussion following
Eq. (\ref{re})]. Internal shocks are, in the wind rest-frame,
``mildly relativistic,'' i.e. characterized by Lorentz factors 
$\gamma_i-1\sim1$. We therefore expect results related to particle
acceleration in sub-relativistic shocks to be valid for the present
scenario. The most restrictive requirement, which rules out the possibility of 
accelerating 
particles to energy $\sim10^{20}$~eV in most astrophysical objects, 
is that
the particle Larmor radius $R_L$ should be smaller than the system size
\cite{Hillas84}. In our scenario we must apply a more stringent requirement,
namely that $R_L$ should be smaller than the largest scale $l$ over which
the magnetic field fluctuates, since otherwise 
Fermi acceleration
may not be efficient. We may estimate $l$ as follows. The comoving time, i.e.
the time measured in the wind rest frame, is $t=r/\Gamma c$. Thus, 
regions separated by a comoving distance larger than $r/\Gamma$ are
causally disconnected, and the wind properties fluctuate over comoving
length scales up to $l\sim r/\Gamma$. We must therefore 
require $R_L<r/\Gamma$.
A somewhat more stringent requirement is related to the wind expansion.
Due to expansion the internal energy is decreasing and therefore
available for proton acceleration (as well as for $\gamma$-ray production) only
over a comoving time $t\sim r/\Gamma c$. The typical Fermi
acceleration time 
is $t_a=f R_L/c\beta^2$, where $\beta c$ is the Alfv\'en velocity and 
$f\sim 1$ \cite{Kulsrud,Hillas84}. 
In our scenario $\beta\simeq1$ leading to the requirement $fR_L<r/\Gamma$.
This condition sets a lower limit to the required
comoving magnetic field strength. This limit may be stated as a radius
independent lower limit to the ratio of magnetic field and electron
energy densities \cite{W95a},
\begin{equation}
\xi_B/\xi_e>0.02 f^2\Gamma_{300}^2 E_{p,20}^2L_{\gamma,52}^{-1},
\label{xiB}
\end{equation}
where $E_p=10^{20}E_{p,20}$~eV is the accelerated proton energy. 

The accelerated proton energy is also limited by energy loss
due to synchrotron radiation and interaction with fireball photons.
As discussed in \S6, the dominant energy loss process is synchrotron cooling. 
The condition that the synchrotron loss time, $t_{sy}=(6\pi m_p^4 c^3/\sigma_T
m_e^2)E^{-1}B^{-2}$, should be smaller than the acceleration time sets
an upper limit to the magnetic field strength. Since the equipartition field
decreases with radius, 
$B_{e.p.}\propto r^{-2}$, the upper limit on the magnetic
field may be satisfied 
simultaneously with (\ref{xiB}) provided
that the internal collisions occur at large enough radius \cite{W95a},
\begin{equation}
r>10^{12}f^2\Gamma_{300}^{-2}E_{p,20}^3{\rm cm}.
\label{rmin}
\end{equation}
Since collisions occur at radius $r\approx\Gamma^2c\Delta t$, the
condition (\ref{rmin}) is equivalent to a lower limit on $\Gamma$
\begin{equation}
\Gamma>130 f^{1/2} E_{20}^{3/4}\Delta t^{-1/4}_{10\rm ms}.
\label{Gmin}
\end{equation}

>From Eqs. (\ref{xiB}) and (\ref{Gmin}), we infer that 
a dissipative ultra-relativistic wind,
with luminosity and variability time implied by GRB observations,
satisfies the constraints necessary to allow the acceleration of protons 
to energy $>10^{20}$~eV, provided that the wind bulk Lorentz factor is
large enough, $\Gamma>100$, and that the
magnetic field is close to equipartition with electrons. The former 
condition, $\Gamma>100$, is remarkably similar to that inferred based on
the $\gamma$-ray spectrum, and $\Gamma\sim300$ is the ``canonical'' 
value assumed in the fireball model. The latter condition, magnetic field
close to equipartition, is commonly assumed to be valid in order to account
for the observed $\gamma$-ray emission [see, e.g., Eq. (\ref{Eg})]. 

Finally, two points should be clarified. First, it has recently been claimed
that ultra-high energy protons would lose most of their energy adiabatically, 
i.e. due to expansion, before they escape the fireball 
\cite{RnM_adiabatic}. This claim is based on the assumptions that
internal shocks, and therefore proton acceleration, occur at $r\sim r_i$ only, 
and that subsequently, $r_i<r<r_e$, the fireball expands
adiabatically. Under these assumptions, protons 
would lose most their energy by the time they escape, at
$r\sim r_e$. However, as emphasized both in this section and
in \S2.1, internal shocks are expected to occur at all radii
$r_i<r<r_e$, and in particular at $r\sim r_e$ during the transition
to self-similar expansion. Thus, proton acceleration to ultra-high energy
is expected to operate at all radii up to $r\sim r_e$, where ultra-high
energy particles escape. 

Second, it has recently been pointed out in \cite{Gallant98} that the
conditions at the external shock driven by the fireball into the ambient
gas are not likely to allow proton acceleration to ultra-high energy. 
Although correct, this observation
is irrelevant for the acceleration in internal shocks,
the scenario considered for UHECR production in GRBs in both \cite{W95a} 
and \cite{Vietri95}.

\subsection{UHECR flux and spectrum}

We have shown in \S3, that 
the present rate at which energy should be produced as $>10^{19}$~eV
protons by cosmological cosmic-ray sources in order to 
produce the observed flux is $E^2 (d\dot n_{CR}/dE)_{z=0}
\approx10^{44}{\rm erg\ Mpc}^{-3}{\rm yr}^{-1}$. 
This rate is, remarkably, comparable to that produced in $\gamma$-rays
by cosmological GRBs. The typical GRB $\gamma$-ray energy,
$E\sim10^{53}$~erg, and the present ($z=0$) GRB rate,
which is estimated to be in the range of $R_{GRB}=1/{\rm Gpc}^3{\rm yr}$
to $R_{GRB}=10/{\rm Gpc}^3{\rm yr}$
(see discussion in \S2.2), implies a present energy generation
rate in the range $\sim10^{44}{\rm erg\ Mpc}^{-3}{\rm yr}^{-1}$
to $\sim10^{45}{\rm erg\ Mpc}^{-3}{\rm yr}^{-1}$.
In addition, since protons are accelerated in the GRB model to high energy
by internal shocks, which in the fireball frame are sub-relativistic,
we may expect a generation spectrum $dN/dE\propto E^{-2}$, consistent
with UHECR observations (see \S3.1).

Thus, GRB fireballs would
produce UHECR flux and spectrum 
consistent with that observed, provided the efficiency
with which the wind kinetic energy is converted to $\gamma$-rays, and
therefore to electron energy, is similar to the efficiency with which it is
converted to proton energy, i.e. to UHECRs \cite{W95a}.
There is, however, one additional point which requires
consideration \cite{W95a}. The energy of the most
energetic cosmic ray detected by the Fly's Eye experiment is in excess of
$2\times10^{20}{\rm eV}$, and that of the most
energetic AGASA event is $\sim2\times10^{20}{\rm eV}$. On a
cosmological scale, the distance traveled by such energetic particles is
small: $<100{\rm Mpc}$ ($50{\rm Mpc}$) for the AGASA (Fly's Eye) event
(e.g., \cite{Aharonian94}). Thus, the detection of these events over a $\sim5
{\rm yr}$ period can be reconciled with the rate of nearby GRBs, $\sim1$
per $100\, {\rm yr}$ out to $100{\rm Mpc}$, only if
there is a large dispersion, $\geq100{\rm yr}$, in the arrival time of protons 
produced in a single burst (This implies that if a direct 
correlation between 
high energy CR events and GRBs, as recently suggested in
\cite{MnU95}, is observed
on a $\sim10{\rm yr}$ time scale, it would be strong evidence {\it against} a 
cosmological GRB origin of UHECRs). 

The required dispersion
is likely to occur due to the combined effects of deflection 
by random magnetic fields and energy dispersion of the particles
\cite{W95a}. 
Consider a proton of energy $E$ propagating through a magnetic field of 
strength $B$ and correlation length
$\lambda$. As it travels a distance $\lambda$, the proton is typically 
deflected by an angle $\alpha\sim\lambda/
R_L$, where $R_L=E/eB$ is the Larmor radius. The
typical deflection angle for propagation over a distance $D$ is
$\theta_s\sim(D/\lambda)^{1/2}\lambda/R_L$. This deflection results in a time
delay, compared to propagation along a straight line, 
\begin{equation}
\tau(E,D)\approx\theta_s^2D/4c\approx
2\times10^5E^{-2}_{20}D_{100}^2\lambda_{\rm Mpc}B_{\rm nG}^2\quad{\rm yr},
\label{delay}
\end{equation}
where $D=100D_{100}{\rm Mpc}$, $\lambda=1\lambda_{\rm Mpc}$~Mpc
and $B=10^{-9}B_{\rm nG}$~G. Here, we have chosen numerical values
corresponding to the current upper bound on the inter-galactic magnetic 
field, $B\lambda^{1/2}\le10^{-9}{\rm G\ Mpc}^{1/2}$ \cite{IGM}.
The random energy loss UHECRs suffer as they propagate, owing to the 
production of pions, implies that 
at any distance from the observer there is some finite spread
in the energies of UHECRs that are observed with a given fixed energy.
For protons with energies
$>10^{20}{\rm eV}$ the fractional RMS energy spread is of order unity
over propagation distances in the range $10-100{\rm Mpc}$ 
(e.g. \cite{Aharonian94}).
Since the time delay is sensitive to the particle energy, this implies that
the spread in arrival time of UHECRs with given observed energy is comparable
to the average time delay at that energy $\tau(E,D)$
(This result has been confirmed by numerical calculations in \cite{Coppi96}).
Thus, the  required time spread, $\tau>100$~yr, is consistent with
the upper bound, $\tau<2\times10^5$~yr, implied by the present
upper bound to the inter-galactic magnetic field.

\section{GRB model predictions for UHECR experiments}

\subsection{The Number and Spectra of Bright Sources}

The initial proton energy, necessary to have an observed energy $E$,
increases with source distance due to propagation energy losses.
The rapid increase of the initial energy after it exceeds, due to
electron-positron production, the threshold for pion production effectively
introduces a cutoff distance, $D_c(E)$, beyond which sources do not contribute
to the flux above $E$. The function $D_c(E)$ is shown in Fig. 2 (taken from
\cite{MnW96}). Since $D_c(E)$ is a decreasing function of $E$, for
a given number density of sources there is a critical energy $E_c$, above which
only one source (on average) contributes to the flux. 
In the GRB model $E_c$ depends on the product of the burst rate $R_{GRB}$
and the time delay. The number of sources contributing, on average, 
to the flux at energy $E$ is \cite{MnW96}
\begin{equation}
N(E) = {4\pi\over 5} R_{GRB}D_c(E)^3 \tau\left[E,D_c(E)\right]\quad,
\label{N}
\end{equation}
and the average intensity resulting from all sources is
\begin{equation}
J(E) = \frac{1}{4\pi}R_{GRB} {d n_p\over dE} D_c(E)\quad,
\label{J}
\end{equation}
where $d n_p/dE$ is the number per unit energy of protons produced on average
by a single burst (this is the formal definition of $D_c(E)$). 
The critical energy $E_c$ is given by
\begin{equation}
{4\pi\over 5} R_{GRB}D_c(E_c)^3 \tau\left[E_c,D_c(E_c)\right]=1\quad.
\label{Ec}
\end{equation}

$E_c$, the energy beyond which a single source contributes on average to
the flux, depends on the unknown properties of 
the intergalactic magnetic field, $\tau\propto B^2\lambda$. 
However, the rapid
decrease of $D_c(E)$ with energy near $10^{20}{\rm eV}$, see Fig. 2,
implies that $E_c$ is only weakly dependent on the value of $B^2\lambda$. 
In The GRB model, the product $R_{GRB}\tau(D=100{\rm Mpc},E=10^{20}{\rm eV})$
is approximately limited to the range $10^{-6}{\rm\ Mpc}^{-3}$ to
$10^{-3}{\rm\ Mpc}^{-3}$ (The lower limit is set by the requirement that 
at least a few GRB sources be present at $D<100$~Mpc, and the upper limit by 
the Faraday rotation bound 
$B\lambda^{1/2}\le10^{-9}{\rm G\ Mpc}^{1/2}$ \cite{IGM} and 
$R_{GRB}\le10/{\rm\ Gpc}^3{\rm yr}$). The corresponding range
of values of $E_c$ is 
$10^{20}{\rm eV}\le E_c<3\times10^{20}{\rm eV}$.

\begin{figure}
\centerline{\psfig{figure=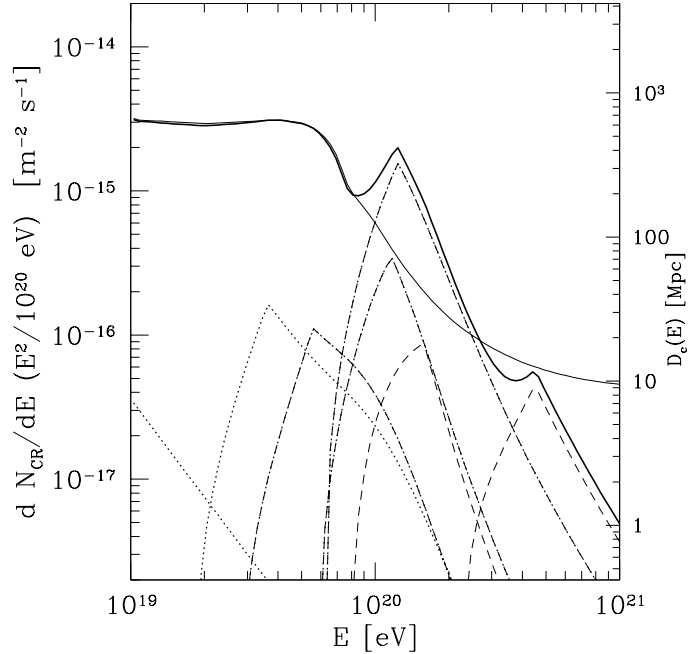,width=3.6in}}
\caption{Results of a Monte-Carlo realization of the bursting sources
model, with $E_c=1.4\times10^{20}$~eV: Thick solid line- overall 
spectrum in the realization;
Thin solid line- average spectrum, this
curve also gives $D_c(E)$;
Dotted lines- spectra of brightest sources at different energies.
}
\label{figNc}
\end{figure}

Fig. 2 presents the flux obtained in one realization of
a Monte-Carlo simulation described by Miralda-Escud\'e \& Waxman 
\cite{MnW96} of the total
number of UHECRs received from GRBs at some fixed time. 
For each
realization the positions (distances from Earth) and
times at which cosmological GRBs occurred were randomly drawn, 
assuming, an intrinsic proton
generation spectrum $dN_p/dE_p \propto E_p^{-2}$, and 
$E_c=1.4\times10^{20}{\rm eV}$. 
Most of the realizations gave an overall spectrum similar to that obtained
in the realization of Fig. \ref{figNc} when the brightest source of this 
realization (dominating at $10^{20}{\rm eV}$) is not included.
At $E < E_c$,
the number of sources contributing to the flux is very large, 
and the overall UHECR flux received at any
given time is near the average (the average flux is that obtained when 
the UHECR emissivity is spatially uniform and time independent).
At $E > E_c$, the flux will generally be much lower than the average,
because there will be no burst within a distance $D_c(E)$ having taken
place sufficiently recently. There is, however, a significant probability
to observe one source with a flux higher than the average.
A source similar to the brightest one in Fig. \ref{figNc}
appears $\sim5\%$ of the time. 

At any fixed time a given burst is observed in UHECRs only over a narrow
range of energy, because if
a burst is currently observed at some energy $E$ then UHECRs of much lower 
energy from this burst have not yet arrived,  while higher energy UHECRs
reached us mostly in the past. As mentioned above, for energies above the 
pion production threshold, 
$E\sim5\times10^{19}{\rm eV}$, the dispersion in arrival times of UHECRs
with fixed observed energy is comparable to the average delay at that
energy. This implies that
the spectral width $\Delta E$ of the source at a given time is of order
the average observed energy, $\Delta E\sim E$.
Thus, bursting UHECR sources should have narrowly peaked energy
spectra,
and the brightest sources should be different at different energies.
For steady state sources, on the other hand, the brightest
source at high energies should also be the brightest one at low
energies, its fractional contribution to the overall flux decreasing to
low energy only as $D_c(E)^{-1}$.
A detailed numerical analysis of the time dependent energy spectrum of 
bursting sources is given in \cite{Sigl97,Lemoine97}.

\subsection{Spectra of Sources at $E<4\times10^{19}{\rm eV}$}
\label{subsec:Blambda}

The detection of UHECRs 
above $10^{20}{\rm eV}$ imply that the brightest sources 
must lie at distances smaller than $100{\rm Mpc}$.
UHECRs with $E\le4\times10^{19}{\rm eV}$
from such bright sources will suffer energy loss only by pair production,
because at $E < 5\times 10^{19}$ eV
the mean-free-path for pion production interaction
(in which the fractional energy loss is $\sim10\%$) is larger than 
$1{\rm Gpc}$. Furthermore, the energy loss due to pair production 
over $100{\rm Mpc}$ propagation is only $\sim5\%$.

In the case where the typical displacement of the UHECRs 
due to deflections by inter-galactic magnetic fields is 
much smaller than the correlation length, $\lambda \gg D\theta_s(D,E)
\simeq D(D/\lambda)^{1/2}\lambda/R_L$,
all the UHECRs that arrive at the
observer are essentially deflected by the same magnetic field structures, 
and the absence of random energy loss during propagation implies that
all rays with a fixed observed energy would reach the observer with exactly
the same direction and time delay. At a fixed time, therefore, the source would
appear mono-energetic and point-like. In reality,
energy loss due to pair production
results in a finite but small spectral and angular width, 
$\Delta E/E\sim\delta\theta/\theta_s\le1\%$ \cite{WnM96}.

In the case where the typical displacement of the UHECRs is 
much larger than the correlation length, $\lambda \ll D\theta_s(D,E)$,
the deflection of different UHECRs arriving at the observer
are essentially independent. Even in the absence of any energy loss there 
are many paths from the source to the observer for UHECRs of fixed energy $E$
that are emitted from the source at an angle 
$\theta\le\theta_s$ relative to the source-observer line of sight. Along
each of the paths, UHECRs are deflected by independent magnetic field 
structures. Thus, the source angular size would be of order $\theta_s$
and the spread in arrival times would be comparable to the characteristic 
delay $\tau$, leading to $\Delta E/E\sim1$ even when there are no random
energy losses. The observed spectral shape of a nearby ($D<100{\rm Mpc}$) 
bursting source of UHECRs at 
$E<4\times10^{19}{\rm eV}$
was derived for the case $\lambda \ll D\theta_s(D,E)$ in 
\cite{WnM96}, and is given by
\begin{equation}
{dN\over dE}\propto \sum\limits_{n=1}^{\infty} (-1)^{n+1}\, n^2\,
\exp\left[ -{2n^2\pi^2 E^2\over E_0^2(t,D)} \right]\quad,
\label{flux}
\end{equation}
where $E_0(t,D)=De(2{B^2\lambda}/3ct)^{1/2}$.  
For this spectrum, the ratio of the 
RMS UHECR energy spread to the average energy is $30\%$

\begin{figure}
\centerline{\psfig{figure=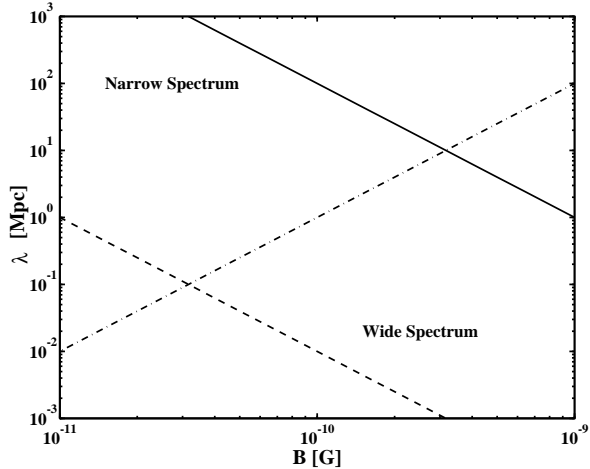,width=3.5in}}
\caption{The line $\theta_s D=\lambda$ for a source at 
$30{\rm Mpc}$ distance 
observed at energy $E\simeq10^{19}{\rm eV}$ (dot-dash line), shown with
the Faraday rotation upper limit $B\lambda^{1/2}
\le10^{-9}{\rm G\ Mpc}^{1/2}$ (solid line), and with the lower limit 
$B\lambda^{1/2}\ge10^{-11}{\rm G\ Mpc}^{1/2}$ required in the GRB model.
}
\label{figBL}
\end{figure}

Fig. \ref{figBL} shows the line $\theta_s D=\lambda$ in the $B-\lambda$ plane,
for a source at a distance $D=30{\rm Mpc}$
observed at energy $E\simeq10^{19}{\rm eV}$.
Since the $\theta_s D=\lambda$ line divides the
allowed region in the plane at $\lambda\sim1{\rm Mpc}$,
measuring the spectral width of bright sources would allow to determine
if the field correlation length is much larger, much smaller, or comparable
to $1{\rm Mpc}$.

\section{High energy Neutrinos}

\subsection{Neutrino production}

\subsubsection{Neutrinos at energies $\sim10^{14}$~eV}

Protons accelerated in the fireball to high energy lose energy through
photo-meson interaction with fireball photons. The decay of charged
pions produced in this interaction, $\pi^+\rightarrow\mu^++\nu_\mu
\rightarrow e^++\nu_e+\overline\nu_\mu+\nu_\mu$, 
results in the production of high energy neutrinos. 
The neutrino spectrum is determined by the observed gamma-ray
spectrum, which is well described by a broken power-law,
$dN_\gamma/dE_\gamma\propto E_\gamma^{-\beta}$ 
with different values of $\beta$ at low and high energy \cite{Band}. The
observed break energy (where $\beta$ changes) is typically 
$E_\gamma^b\sim1{\rm MeV}$, 
with $\beta\simeq1$ at energies below the break and $\beta\simeq2$ 
above the break. The interaction of protons accelerated to a power-law
distribution, $dN_p/dE_p\propto E_p^{-2}$, 
with GRB photons results in a broken power law
neutrino spectrum \cite{WnB97}, $dN_\nu/dE_\nu\propto E_\nu^{-\beta}$ with
$\beta=1$ for $E_\nu<E_\nu^b$, and $\beta=2$ for $E_\nu>E_\nu^b$. 
The neutrino break energy $E_\nu^b$ is fixed by the threshold energy
of protons for photo-production in interaction with the dominant $\sim1$~MeV
photons in the GRB,
\begin{equation}
E_\nu^b\approx5\times10^{14}\Gamma_{300}^2(E_\gamma^b/1{\rm MeV})^{-1}{\rm eV}.
\label{Enu}
\end{equation}

The normalization of the flux is determined by the
efficiency of pion production.
As shown in \cite{WnB97}, the fraction of energy lost to pion production
by protons producing the neutrino flux above the break, $E^b_\nu$, is 
essentially independent of energy and is given by
\begin{equation}
f_\pi\approx0.2{L_{\gamma,52}\over
(E_\gamma^b/1{\rm MeV})\Gamma_{300}^4 \Delta t_{10\rm  ms}}.
\label{fpi}
\end{equation}
Thus, acceleration of protons to high energy in internal fireball
shocks would lead to conversion of a significant fraction of
proton energy to high energy neutrinos.

If GRBs are the sources of UHECRS, 
then using Eq. (\ref{fpi}) and the UHECR generation rate implied by 
observations, the expected GRB neutrino flux is \cite{WnB99}
\begin{eqnarray}
E_\nu^2\Phi_{\nu_\mu}&&\approx E_\nu^2\Phi_{\bar\nu_\mu}
\approx E_\nu^2\Phi_{\nu_e}\cr
&&\approx 1.5\times10^{-9}\left({f_\pi\over0.2}\right)\min\{1,E_\nu/E^b_\nu\}
{\rm GeV\,cm}^{-2}{\rm s}^{-1}{\rm sr}^{-1}.
\label{JGRB}
\end{eqnarray}

The GRB neutrino flux can be estimated directly from
the observed gamma-ray fluence. The Burst and Transient Source Experiment 
(BATSE) measures the GRB fluence $F_\gamma$ over two decades of photon energy, 
$\sim0.02$MeV to $\sim2$MeV, corresponding to a decade of radiating
electron energy (the electron 
synchrotron frequency is proportional to the square of
the electron Lorentz factor). If electrons carry a fraction $f_e$ of
the energy carried by protons, then the muon neutrino fluence of a single
burst is $E_\nu^2dN_\nu/dE_\nu\approx0.25(f_\pi/f_e)F_\gamma/\ln(10)$. 
The average neutrino flux per unit time and solid angle is obtained by
multiplying the single burst fluence with the GRB rate per solid angle,
$\approx10^3$ bursts per year over $4\pi$~sr. Using the average burst
fluence $F_\gamma=10^{-5}{\rm erg/cm}^2$, we obtain
a muon neutrino flux $E_\nu^2\Phi_\nu\approx3\times10^{-9}(f_\pi/f_e)
{\rm GeV/cm}^2{\rm s\,sr}$. Thus, the 
neutrino flux estimated directly from the gamma-ray fluence agrees with
the estimate (\ref{JGRB}) based on the cosmic-ray production rate. 

\subsubsection{Neutrinos at high energy $>10^{16}$~eV}

The neutrino spectrum (\ref{JGRB}) is
modified at high energy, where neutrinos are produced by the decay
of muons and pions whose life time $\tau_{\mu,\pi}$
exceeds the characteristic time for
energy loss due to adiabatic expansion and synchrotron emission 
\cite{WnB97,RnM98,WnB99}.
The synchrotron loss time is determined by the energy density of the
magnetic field in the wind rest frame.
For the characteristic parameters of a GRB wind, the muon energy for which
the adiabatic energy loss time equals the muon life time,
$E^a_\mu$, is comparable to the energy $E^s_\mu$ 
at which the life time equals 
the synchrotron loss time, $\tau^s_\mu$. For pions, $E^a_\pi>E^s_\pi$. This,
and the fact that the adiabatic
loss time is independent of energy and the synchrotron loss time is
inversely proportional to energy, imply that
synchrotron losses are the dominant effect
suppressing the flux at high energy. The energy above which synchrotron
losses suppress the neutrino flux is
\begin{equation}
{E^s_{\nu_\mu(\bar\nu_\mu,\nu_e)}\over E^b_\nu}
\approx(\xi_B L_{\gamma,52}/\xi_e)^{-1/2}\Gamma_{300}^2
\Delta t_{10\rm ms}(E^b_\gamma/1{\rm MeV})\times
\cases{10,&for $\bar\nu_\mu$, $\nu_e$;\cr 100,&for $\nu_\mu$ .\cr}
\label{nu_sync}
\end{equation}

We note, that the results presented above were derived using the 
``$\Delta$-approximation,'' i.e.
assuming that photo-meson interactions are dominated by the contribution of
the $\Delta$-resonance.
It has recently been shown \cite{Muecke98}, that for photon spectra harder
than $dN_\gamma/dE_\gamma\propto E^{-2}_\gamma$, the contribution of 
non-resonant interactions may be important. Since in order to interact with
the hard part of the photon spectrum, $E_\gamma<E_\gamma^b$, the proton energy
must exceed the energy at which neutrinos of energy $E_\nu^b$ are
produced, significant modification of the $\Delta$-approximation results
is expected only for $E_\nu\gg E_\nu^b$, where the neutrino flux is 
strongly suppressed by synchrotron losses.

So far, we have discussed neutrino production in internal shocks
due to variability on the shortest time scale, $\Delta t\sim10$~ms.
Internal collisions due to variability on longer time scales,
$\Delta t<\delta t<T$, are less efficient in 
producing neutrinos, $f_\pi\propto\delta t^{-1}$ [cf. Eq. (\ref{fpi})], 
since the radiation energy 
density is lower at larger collision radii. However, at larger radii 
synchrotron losses cut off
the spectrum at higher energy, $E^s(\delta t)\propto\delta t$ [cf. Eq. 
(\ref{nu_sync})]. Collisions at large radii therefore
result in extension of the neutrino
spectrum of Eq. (\ref{JGRB}) to higher energy, beyond the cutoff energy 
Eq. (\ref{nu_sync}), 
\begin{equation}
E^2_\nu\Phi_\nu\propto E_\nu^{-1},\quad E_\nu>E^s_\nu.
\label{high_nu}
\end{equation}

The neutrino flux from GRBs is small above $10^{19}$eV, 
and a neutrino flux comparable to the $\gamma$-ray flux
is expected only below $\sim10^{17}$eV, in agreement with the
results of ref. \cite{RnM98}.
Our result is not in agreement, however, with that of ref. \cite{Vietri_nu19}, 
where
a much higher flux at $\sim10^{19}$eV is obtained based on the equations
of ref. \cite{WnB97}, which are the same equations as used 
here\footnote{The parameters chosen in \cite{Vietri_nu19} 
are $L_\gamma=10^{50}{\rm
erg/s}$, $\Delta t=10$s, and $\Gamma=100$. Using equation (4) of ref.
\cite{WnB97}, which is the same as Eq. (\ref{fpi}) of the present paper,
we obtain for these parameters $f_\pi=1.6\times10^{-4}$, while the author
of \cite{Vietri_nu19} obtains, using the same equation, $f_\pi=0.03$.}.
Finally, we note that, contrary to the claim in \cite{RnM98},
there is no contradiction between production of high-energy
protons above $\sim3\times10^{20}$eV and a break in the neutrino
spectrum at $\sim10^{16}$eV [cf. Eqs. (\ref{xiB},\ref{Gmin},\ref{nu_sync})].

\subsection{Implications}

The high energy neutrinos predicted in the dissipative wind model of GRBs
may be observed by detecting the Cerenkov light emitted by high energy muons
produced by neutrino interactions below a detector on the surface of the
Earth (see \cite{Gaisser} for a recent review). 
The probability $P_{\nu\mu}$ that a neutrino
would produce a high energy muon in the detector is approximately given by 
the ratio of the high energy muon range to the neutrino mean free path.
At the high energy we are considering, 
$P_{\nu\mu}\simeq10^{-6}(\epsilon_\nu/1{\rm TeV})$ \cite{Gaisser}.
Using Eq. (\ref{JGRB}), this implies 
a detection rate of $\sim20$
neutrino induced muons per year for a $1\,{\rm km}^2$ detector 
(over $4\pi$~sr). As discussed in \cite{WnB97}, one may look
for neutrino events in angular coincidence, on degree scale, 
and temporal coincidence, on time scale of seconds, with GRBs. 
Several authors \cite{nu_attenuation} have recently 
emphasized the effect on muon detection rate of neutrino absorption in 
the Earth. This effect is not large for GRB neutrinos, since most of the 
signal comes from neutrinos of energy $\sim10^{14}$~eV. At this energy,
the flux of upward moving muons is reduced due to absorption by
$36\%$ \cite{nu_attenuation}, 
and the total ($4\pi$~sr) flux is reduced by only $18\%$.

Detection of neutrinos from GRBs could be used to
test the simultaneity of
neutrino and photon arrival to an accuracy of $\sim1{\rm\ s}$
($\sim1{\rm\ ms}$ for short bursts), checking the assumption of 
special relativity
that photons and neutrinos have the same limiting speed
[The time delay for neutrino of energy $10^{14}{\rm eV}$ 
with mass $m_\nu$ traveling $100{\rm\ Mpc}$ is only 
$\sim10^{-11}(m_\nu/10{\rm\ eV})^2{\rm s}$].
These observations would also test the weak
equivalence principle, according to which photons and neutrinos should
suffer the same time delay as they pass through a gravitational potential.
With $1{\rm\ s}$ accuracy, a burst at $100{\rm\ Mpc}$ would reveal
a fractional difference in limiting speed 
of $10^{-16}$, and a fractional difference in gravitational time delay 
of order $10^{-6}$ (considering the Galactic potential alone).
Previous applications of these ideas to supernova 1987A 
(see \cite{John_book} for review), where simultaneity could be checked
only to an accuracy of order several hours, yielded much weaker upper
limits: of order $10^{-8}$ and $10^{-2}$ for fractional differences in the 
limiting speed \cite{c} and time delay \cite{WEP} respectively.

The model discussed above predicts the production of high energy
muon and electron neutrinos
with a 2:1 ratio. If vacuum neutrino oscillations occur in nature, 
then neutrinos that get here should be almost equally distributed between
flavors for which the mixing is strong.
In fact, if the atmospheric neutrino anomaly has the explanation it is
usually given, oscillation to $\nu_\tau$'s with mass $\sim0.1{\rm\ eV}$
\cite{atmo}, then
one should detect equal numbers of $\nu_\mu$'s and $\nu_\tau$'s. 
Upgoing $\tau$'s, rather than $\mu$'s, would be a
distinctive signature of such oscillations. 
Since $\nu_\tau$'s are not expected to be produced in the fireball, looking
for $\tau$'s would be an ``appearance experiment''
($\nu_\tau$'s may be produced by photo-production of charmed mesons; 
However, the high photon threshold, $\sim50{\rm GeV}$, and
low cross-section, $\sim1\mu{\rm b}$ \cite{charm}, for such reactions imply 
that the ratio of charmed meson to pion production is $\sim10^{-4}$).
To allow flavor change, the difference in squared neutrino masses, 
$\Delta m^2$, should exceed a minimum value
proportional to the ratio of source
distance and neutrino energy \cite{John_book}. A burst at $100{\rm\ Mpc}$ 
producing $10^{14}{\rm eV}$ neutrinos can test for $\Delta m^2\ge10^{-16}
{\rm eV}^2$, 5 orders of magnitude more sensitive than solar neutrinos.
Note, that due to the finite pion life time, flavor mixing would be caused by
de-coherence, rather than by real oscillations, for neutrinos with masses
$>0.1{\rm eV}$.

\section{Summary}

Afterglow observations confirmed the cosmological origin of GRBs and provide
support for the fireball model (\S1,\S2). In this model, observed
radiation is produced by synchrotron emission of shock accelerated electrons.
We have shown, that in the region where electrons are accelerated protons
can be accelerated to ultra-high energy (\S4.1). 
Acceleration to $>10^{20}$~eV is possible
provided that the fireball bulk Lorentz factor is
large enough, $\Gamma>100$, and that the
magnetic field is close to equipartition with electrons. The former 
condition, $\Gamma>100$, is remarkably similar to that inferred based on
$\gamma$-ray spectra, and $\Gamma\sim300$ is the ``canonical'' 
value assumed in the fireball model. The latter condition is commonly 
assumed to be valid in order to account
for observed $\gamma$-ray emission. 

Observed UHECR flux and spectrum are consistent with a model where
UHECRs are protons accelerated to high energy in GRB fireballs,
provided the efficiency 
with which fireball kinetic energy is converted to $\gamma$-rays, and
therefore to accelerated 
electron energy, is similar to the efficiency with which it is
converted to accelerated proton energy (\S3,\S4.2).

The GRB model for UHECR production
has several unique predictions (\S5). 
In particular, a critical energy 
is predicted to exist, $10^{20}{\rm eV}\le E_c<3\times10^{20}{\rm eV}$,
above which a few sources produce most of the UHECR flux, and the 
observed spectra of these sources is predicted to be 
narrow, $\Delta E/E\sim1$: the bright sources
at high energy should be absent in UHECRs of much
lower energy, since particles take longer to arrive the lower their
energy.
Recently, the AGASA experiment reported the presence
of 3 pairs of UHECRs with angular separations (within each pair) 
$\le2.5^\circ$, roughly consistent with the measurement error,
among a total of 36 UHECRs with $E\ge4\times10^{19}{\rm eV}$ 
\cite{AGASA_pairs}. 
The two highest energy AGASA events were in these pairs.
Given the total solid angle observed by the experiment, $\sim2\pi{\rm sr}$,
the probability to have found 3 pairs by chance is $\sim3\%$; and, given that
three pairs were found, the probability that the two highest energy events are
among the three pairs by chance is 2.4\%. Therefore, this observation favors
the bursting source model, although more data are needed to confirm it.
Testing the above predictions of the fireball model for UHECR production
would require an exposure 10 times larger than that of present
experiments. Such increase is expected to be provided by the planned
HiRes \cite{HiRes} and Auger \cite{Auger} detectors.

A natural consequence of proton acceleration in fireball shocks
is the conversion of a large
fraction, $\ge10\%$, of the fireball energy 
to a burst of $\sim10^{14}{\rm eV}$ neutrinos by photo-meson production
(\S6.1). 
Large area, $\sim1{\rm km}^2$, high-energy neutrino telescopes,
which are being constructed to detect 
cosmologically distant neutrino sources, would
observe several tens of events per year correlated with GRBs, and
test for neutrino properties (e.g. flavor oscillations,
for which upward moving $\tau$'s would be a unique signature, and coupling
to gravity) with an accuracy many orders 
of magnitude better than is currently possible (\S6.2).

\end{document}